\documentclass[11pt,a4paper]{article}
\usepackage{amssymb,amscd}
\usepackage{epsfig} 
\newcommand{\rem}[1]{}

\title{On the non-integrability of a fifth order
equation with 
integrable  
two-body dynamics}

\author{ 
D.D. Holm\thanks{Theoretical Division and Center for
Nonlinear Studies, Los Alamos National Laboratory, Los
Alamos, NM 87545, USA. E-mail:dholm@lanl.gov}
\& A.N.W. Hone\thanks{Institute of Mathematics \&
Statistics,
University of Kent, Canterbury CT2 7NF, UK.
E-mail:anwh@ukc.ac.uk}
}

\begin{document}
\renewcommand{\theequation}{\arabic{section}.\arabic{equation}}
\newcommand{\beq}{\begin{equation}}
\newcommand{\eeq}{\end{equation}}
\newcommand{\bea}{\begin{eqnarray}}
\newcommand{\eea}{\end{eqnarray}}
\maketitle

\begin{abstract} 
We consider the fifth order 
partial differential equation (PDE) 
$$ 
u_{4x,t}-5u_{xxt}+4u_t+uu_{5x}+2u_xu_{4x}-5uu_{3x}-10u_xu_{xx}+12uu_x=0,
$$ 
which is a generalization 
of the integrable Camassa-Holm equation. 
The fifth order PDE has exact solutions
in terms of 
an arbitrary number of superposed pulsons, 
with geodesic Hamiltonian dynamics that is  
known to be integrable in the two-body case $N=2$.  
Numerical simulations show that the pulsons are stable, 
dominate the initial value problem and scatter elastically. 
These characteristics are reminiscent of solitons in 
integrable systems. However, after demonstrating 
the non-existence of a suitable Lagrangian or
bi-Hamiltonian 
structure, and obtaining  negative results from
Painlev\'{e} 
analysis and the Wahlquist-Estabrook method, 
we assert that the fifth order PDE is not integrable. 
\end{abstract}

\section{Introduction}

This note is concerned with the fifth order 
partial differential equation (PDE) 
\beq 
u_{4x,t}-5u_{xxt}+4u_t+uu_{5x}+2u_xu_{4x}-5uu_{3x}-10u_xu_{xx}+12uu_x=0.
\label{eq:fifth} 
\eeq 
One reason for our interest in this equation is that 
it admits exact solutions of the form 
\beq 
u=\sum_{j=1}^Np_j(t)(2e^{-|x-q_j(t)|}-e^{-2|x-q_j(t)|}),
\label{eq:pulsons} 
\eeq 
where $p_j,q_j$ satisfy the canonical Hamiltonian
dynamics generated by
\beq 
H_N=\frac{1}{2}\sum_{j,k=1}^Np_jp_k
(2e^{-|q_j-q_k|}-e^{-2|q_j-q_k|}).  
\label{eq:ham} 
\eeq 
Following \cite{fringer}, we refer to such solutions
as ``pulsons.''  
The equations for the $N$-body pulson dynamics are
equivalent to 
geodesic flow on an $N$-dimensional space with
coordinates $q_j$ 
and co-metric 
$$ 
g^{jk}=g(q_j-q_k), \qquad 
g(x)=2e^{-|x|}-e^{-2|x|}.  
$$
The pulsons (\ref{eq:pulsons})
are weak solutions with discontinuous second
derivatives 
at isolated points.


\begin{figure}[ht!]
\centerline{
\scalebox{0.45}{\includegraphics{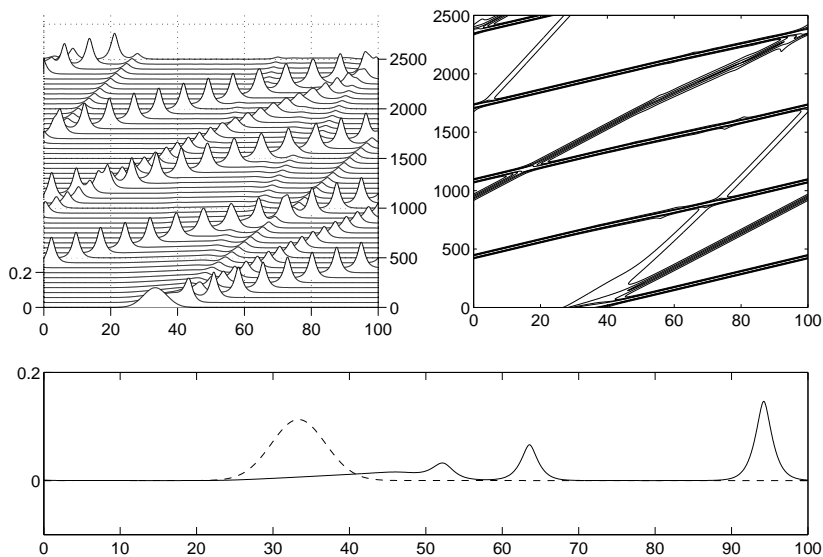}}
}
\caption{Pulson solutions (\ref{eq:pulsons}) of
equation
  (\ref{eq:fifth}) emerge from a Gaussian of unit
  area and width $\sigma = 5$ centered about $x = 33$
on a periodic domain
  of length $L = 100$.  The fastest pulson crosses the
domain four times
  and collides elastically with the slower ones.}
\label{Gauss_InitialCond-waterfall/contour_fig}
\end{figure}

The PDE (\ref{eq:fifth}) is one of a family of
integral partial 
differential equations considered in \cite{fringer},
given by  
\beq 
m_t+um_x+2u_xm=0, \qquad u=g*m, \label{eq:intpde}  
\eeq 
where $u(x,t)$ is defined in terms of $m(x,t)$ by the 
convolution integral 
$$ 
g*m:=\int_{-\infty}^\infty g(x-y)m(y,t)\, dy. 
$$ 
The integral kernel $g(x)$ is taken to be an even
function, 
and for any $g$ the equation (\ref{eq:intpde}) has the
Lie-Poisson Hamiltonian form 
\beq 
m_t=-(m\partial_x+\partial_x m)\frac{\delta H}{\delta
m}  
\label{eq:lph} 
\eeq  
where 
\beq 
H=\frac{1}{2}\int m\, g*m\, dx=\frac{1}{2}\int mu \,
dx. 
\label{eq:pdeham} 
\eeq 
Any equation in this family admits pulson solutions 
$$ 
u(x,t)=\sum_{j=1}^N p_j(t)g(x-q_j(t)) 
$$ 
for arbitrary $N$, with $p_j,q_j$ satisfying the
canonical 
Hamilton's equations 
\beq 
\frac{dp_j}{dt}=-\frac{\partial H_N}{\partial q_j}= 
-p_j\sum_{k=1}^N p_k g'(q_j-q_k), 
\qquad  \frac{dq_j}{dt}=\frac{\partial H_N}{\partial
p_j}= 
\sum_{k=1}^N p_k g(q_j-q_k). 
\label{eq:geo} 
\eeq 
generated by the Hamiltonian  
$$ 
H_N=\frac{1}{2}\sum_{j,k}p_jp_k\, g(q_j-q_k). 
$$  
The equations (\ref{eq:geo}) correspond to geodesic
motion on  
a manifold with co-metric $g^{jk}=g(q_j-q_k)$. 
A significant result of \cite{fringer} is that
the two-body dynamics ($N=2$) is integrable 
for any choice of kernel $g$, 
and numerical calculations show that this elastic
two-pulson 
scattering dominates the initial value problem. 


\begin{figure}[ht!]
\centerline{
\scalebox{0.45}{\includegraphics{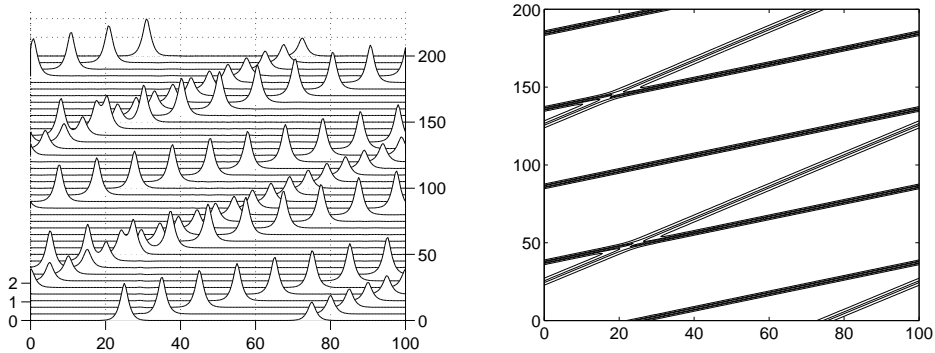}}
}
\caption{Two rear-end collisions of pulson solutions
(\ref{eq:pulsons}) of
  equation (\ref{eq:fifth}).   The initial positions
  are $x=25$ and $x=75$. 
  The faster pulson moves at twice the speed of the
  slower one.  For this ratio of speeds,
  both collisions result in a phase shift to the right
for the faster
  space-time trajectory, but no phase shift for the
slower one.}
\label{rear_collision-waterfall/contour_fig}
\end{figure}
 

Three special cases are isolated in \cite{fringer},
namely 
(up to suitable scaling)  
\begin{itemize} 
\item $g(x)=\delta(x)$ - Riemann shocks,  
\item $g(x)=1-|x|$, $|x|<1$ - compactons,  
\item $g(x)=e^{-|x|}$ - peakons.  
\end{itemize}  
For each of these cases both the integral PDE 
(\ref{eq:intpde}) and the 
corresponding finite-dimensional 
system (\ref{eq:geo}) (for any $N$) are integrable. Of
most 
relevance here is the third case, where
$g(x)=e^{-|x|}$, which  
is the (scaled) Green's 
function for the Helmholtz operator, satisfying the
identity  
$$ 
(1-\partial_x^2)g(x)=2\delta(x).  
$$ 
In that case after rescaling we may take   
\beq 
m=u-u_{xx}, 
\label{eq:mdefpeak} 
\eeq 
and the equation (\ref{eq:intpde}) 
is just  a PDE for $u(x,t)$, namely    
\beq 
u_t-u_{xxt}-uu_{3x}- 2u_xu_{xx}+3uu_x=0, 
\label{eq:ch} 
\eeq 
which is the dispersionless form of the 
integrable Camassa-Holm equation 
for shallow water waves \cite{ch, ch2}.  
For Camassa-Holm the pulson solutions take the form of
peakons or peaked solitons, i.e. 
\beq 
u(x,t)=\sum_{j=1}^N p_j(t)e^{-|x-q_j(t)|}. 
\label{eq:peakons} 
\eeq 

The fifth order equation arises from a different
choice of 
Green's function. Using the identity  
\beq 
(4-\partial_x^2)(1-\partial_x^2)g(x)=12\delta(x),
\qquad 
g(x)=2e^{-|x|}-e^{-2|x|} 
\label{eq:green} 
\eeq 
we find that (after suitable scaling) this choice of 
$g$ yields 
\beq 
m=u_{4x}-5u_{xx}+4u,
\label{eq:mdeffif} 
\eeq 
and then the equation (\ref{eq:intpde}) becomes 
the fifth order equation (\ref{eq:fifth}). Thus  the
PDE 
(\ref{eq:fifth}) should be regarded as a natural
higher 
order generalization of the Camassa-Holm equation.   

\section{Hamiltonian and Lagrangian considerations} 

\setcounter{equation}{0}

In a 
forthcoming article \cite{dhh} 
we discuss a more general family 
of integral PDEs of the form 
\beq 
m_t+um_x+bu_x m =0, \qquad u=g*m,  
\label{eq:bfamily} 
\eeq 
where $b$ is an arbitrary parameter; (\ref{eq:intpde})
corresponds to the particular case $b=2$, and
numerical 
results have recently been established for different
$b$ values 
in \cite{hs}. In the case 
of the peakon kernel $g=e^{-|x|}$, with $m$ given by 
(\ref{eq:mdefpeak}), the equations in this class were 
tested by the method of asymptotic integrability
\cite{dega}, 
and only the cases $b=2,3$ were isolated as
potentially integrable. 
For $b=2$ the integrability of the 
Camassa-Holm equation by inverse scattering 
was already known \cite{ch, ch2},   
but for the new equation 
\beq 
u_t-u_{xxt}-uu_{3x}- 3u_xu_{xx}+4uu_x=0,
\label{eq:td}
\eeq
with $b=3$ the integrability was proved in
\cite{needs} by the 
construction of the Lax pair.   The two integrable
cases 
$b=2,3$ were also found recently via the perturbative
symmetry 
approach \cite{mik}. For any $b\neq -1$, the peakon family 
(\ref{eq:bfamily}) with $g=e^{-|x|}$ arises as the dispersionless 
limit at the quadratic order in the asymptotic expansion for 
shallow water waves \cite{dgh2}.   
 
Another motivation for our interest 
in the fifth order equation (\ref{eq:fifth}) is that
it 
is expressed naturally in terms of the quantity $m$ 
(\ref{eq:mdeffif}),   
given by the product of two Helmholtz operators acting
on $u$.      
Such a product appears in the fifth order operator 
$$
B_0=\partial_x(4-\partial_x^2)(1-\partial_x^2),  
$$ 
which 
we found \cite{needs} to provide 
the first Hamiltonian structure 
for the  
the new integrable equation (\ref{eq:td}). This led us
to the conjecture that the operator $B_0$ should
appear  
naturally in the theory of 
higher order integrable equations such as
(\ref{eq:fifth}).   

All the equations of the form (\ref{eq:intpde}) 
have the Lie-Poisson Hamiltonian structure given by 
(\ref{eq:lph}), but for integrability we expect a 
bi-Hamiltonian structure. In the case of the
Camassa-Holm 
equation (\ref{eq:ch}) 
there are two ways to derive a second Hamiltonian
structure. 
The first is by inspection using a conservation law,
noting that 
(\ref{eq:ch}) may be written as 
\beq 
m_t=\left(uu_{xx}+\frac{1}{2}u_x^2-\frac{3}{2}u^2\right)_x
=\partial_x \frac{\delta \tilde{H}}{\delta u} 
=\partial_x(1-\partial_x^2)\frac{\delta
\tilde{H}}{\delta m} 
\label{eq:biham} 
\eeq 
for 
\beq 
\tilde{H}=-\frac{1}{2}\int (uu_x^2+u^3)\, dx. 
\label{eq:nham} 
\eeq 
The identity (\ref{eq:biham}) gives the second
Hamiltonian structure 
for Camassa-Holm, and $m\partial_x+\partial_x m$,  
$\partial_x(1-\partial_x^2)$ constitute a 
compatible bi-Hamiltonian pair. 
  
Similarly $\int m \, dx$ is conserved for
(\ref{eq:fifth}), 
with $m$ given by (\ref{eq:mdeffif}). 
The conservation law 
is explicitly      
\begin{equation} 
(u_{4x}-5u_{xx}+4u)_t=-\left(uu_{4x}+u_xu_{3x} 
-\frac{1}{2}u_{xx}^2-5uu_{xx}-\frac{5}{2}u_x^2
+6u^2\right)_x 
=:\mathcal{F}_x. 
\label{eq:fif} 
\end{equation} 
By analogy with the Camassa-Holm equation, 
this would suggest that a suitable constant
coefficient 
Hamiltonian operator might be  
$\partial_x(4-\partial_x^2)(1-\partial_x^2)$ (which we
know to be  
a Hamiltonian operator for the new equation
(\ref{eq:td})). 
This would require the 
right hand side of (\ref{eq:fif}) to take the form  
$$ 
\partial_x\frac{\delta K}{\delta u}= 
\partial_x(4-\partial_x^2)(1-\partial_x^2)\frac{\delta
K}{\delta m}. 
$$ 
However, for the flux of (\ref{eq:fif}) we find  
$$
\mathcal{F}\neq \frac{\delta K}{\delta u} 
$$ 
for any local density functional $K$ of $u$,  and we  
suppose that the 
operators $\partial_x(4-\partial_x^2)(1-\partial_x^2)$
and $m\partial_x+\partial_x m$ must be incompatible. 
 
The second way to derive the Hamiltonian structure
(\ref{eq:biham}) 
for Camassa-Holm is via the action (integral of Lagrangian density)  
$$ 
S=\int\int \mathcal{L}[\phi]\, dx\,dt:=\int\int 
\frac{1}{2}\left( 
\phi_x\phi_t-\phi_{3x}\phi_t+\phi_x\phi_{xx}^2+\phi_x^3\right)
\, dx\,dt,  
$$   
for $u=\phi_x$. A Legendre transformation yields the
conjugate 
momentum 
$$ 
\frac{\partial \mathcal{L}}{\partial \phi_t} 
=\frac{1}{2}(\phi_x-\phi_{3x})=
\frac{m}{2}, 
$$ 
and the same Hamiltonian as (\ref{eq:nham}) above,
i.e. 
$$ 
\tilde{H}=\int\left(\frac{1}{2}m\phi_t-\mathcal{L}\right)\,
dx. 
$$ 
Trying the same approach for (\ref{eq:fifth}), we set
$u=\phi_x$ 
and rewrite it as 
\beq 
\phi_{5x,t}-5\phi_{3x,t}+4\phi_{xt}  
+\phi_x\phi_{6x}+2\phi_{xx}\phi_{5x} 
-5\phi_x\phi_{4x}-10\phi_{xx}\phi_{3x}+12\phi_x\phi_{xx}=0.
\label{eq:qlag} 
\eeq 
However, the equation (\ref{eq:qlag}) cannot be
derived from 
a local Lagrangian density $\mathcal{L}[\phi]$ due to
the 
presence of the terms
$\phi_x\phi_{6x}+2\phi_{xx}\phi_{5x}$.       

The first nonlocal Hamiltonian structure for the
Camassa-Holm 
equation is obtained by applying the recursion
operator to 
$m\partial_x+\partial_x m$. This means that
(\ref{eq:ch}) 
can be written in the Hamiltonian form 
$$ 
m_t= 
(m\partial_x+\partial_x m)(\partial_x^3-\partial_x
)^{-1} 
( m\partial_x+\partial_x m) 
\frac{\delta \hat{H}}{\delta m}, \qquad \hat{H}=\int
m\, dx. 
$$   
With the same $\hat{H}$, 
the analogous identity for (\ref{eq:fifth}) is 
\beq 
m_t=B\frac{\delta \hat{H}}{\delta m}\equiv 
(m\partial_x+\partial_x m)
(\partial_x^5-5\partial_x^3+4\partial_x )^{-1} 
( m\partial_x+\partial_x m)
\frac{\delta \hat{H}}{\delta m}, 
\label{eq:nonloc} 
\eeq   
but from the above considerations we would 
expect that the formal 
nonlocal operator $B$ on the right hand 
side of (\ref{eq:nonloc}) is not 
Hamiltonian, and indeed using the functional equations
derived  
in \cite{dhh} it is possible to show that it fails to
satisfy 
the Jacobi identity.   

\section{Reciprocal transformation and Painlev\'{e}
analysis} 

\setcounter{equation}{0}

Having failed to find the sort of Lagrangian or
bi-Hamiltonian 
structure for (\ref{eq:fifth}) that we would
reasonably expect, 
we proceed to see what Painlev\'{e} analysis can tell
us about 
this fifth order equation. However, we note that both
the 
Camassa-Holm equation (\ref{eq:ch}) and the new
equation 
(\ref{eq:td}) provide examples of the weak
Painlev\'{e} 
property \cite{weak}, with algebraic branching in the 
solutions. For these equations we have found it
convenient  
to use  
reciprocal transformations (see \cite{rogers1} for 
definitions), which transform to equations with pole
singularities, 
and indeed in \cite{needs} this was the key 
to our discovery of the Lax pair for (\ref{eq:td}).
Hodograph 
transformations of this kind have been used before to
remove 
branching from classes of evolution equations
\cite{hodo, hone}, but 
here we are dealing with non-evolutionary PDEs. 

To make the results of our analysis more general, we
will 
consider the whole class of equations 
\beq 
m_t+um_x+bu_x m =0, \qquad m=u_{4x}-5u_{xx}+4u, \qquad
b\neq 0,  
\label{eq:fifbfamily}
\eeq
for arbitrary nonzero $b$,  which is the particular
family of 
equations (\ref{eq:bfamily}) corresponding to the 
integral kernel (\ref{eq:green}), and includes 
(\ref{eq:fifth}) in the special case $b=2$. Each
equation in the 
class (\ref{eq:fifbfamily}) has the conservation law 
$$ 
(m^{1/b} )_t =-(m^{1/b}u)_x, 
$$ 
and so introducing a new dependent variable $p$
according to 
\beq 
p^b=-m 
\label{eq:pdef} 
\eeq 
means that we may consistently define a reciprocal  
transformation to new independent variables $X,T$
given 
by 
\beq 
dX=p\, dx-pu\,dt, \qquad dT=dt.
\label{eq:recip}
\eeq
Transforming the derivatives we have the new
conservation law
\beq
(p^{-1})_T=u_X. \label{eq:ncon}
\eeq
Rewriting the relation 
(\ref{eq:mdeffif}) in terms of $\partial_X$
and using (\ref{eq:ncon}) 
to eliminate derivatives of $u$ we
obtain the identity
\beq
u=\frac{1}{4}\left(5-(p\partial_X)^2\right)\cdot
(p\partial_X)\cdot p(p^{-1})_T-\frac{p^b}{4}, 
\label{eq:uid}
\eeq
which means that (\ref{eq:ncon}) can be written as
an equation for $p$ alone, i.e. 
\beq 
(p^{-1})_T=\left( 
\frac{1}{4}\left((p\partial_X)^2-5\right)p(\log
p)_{XT}- 
\frac{p^b}{4}\right)_X. 
\label{eq:refif} 
\eeq 

The fifth order equation (\ref{eq:refif}) is the
reciprocal 
transform of (\ref{eq:fifbfamily}).                   
Rather than carrying out the full Painlev\'{e} test
for the 
transformed equation, it is sufficient for our
purposes 
to follow \cite{ars} and apply the Painlev\'{e} test
for   
ODEs to the travelling wave reduction of
(\ref{eq:refif}).  
Hence we set $p=p(z)$, $z=X-cT$ and the resulting
fifth order 
ODE may be integrated twice 
to get the third order ODE 
\beq  
\frac{5}{8}\left(\frac{p'}{p}\right)^2-\frac{1}{4}\left(p'p'''
-\frac{1}{2}(p'')^2-\frac{(p')^2p''}{p}+
\frac{1}{2}\frac{(p')^4}{p^2}\right) 
-\frac{1}{2p^2}=\frac{c^{-1}p^{b-1}}{4(b-1)}+\frac{d}{p}+e,
\label{eq:ode} 
\eeq 
$b\neq 1$, where $d,e$ are arbitrary constants and $c$
is the wave 
speed, 
with prime denoting $d/dz$. For $b=1$ there is a $\log
p$ term 
on the right hand side, and so this case has
logarithmic 
branching and is immediately excluded  
by the Painlev\'{e} test. Similarly, because of the
$p^{b-1}$ 
term  all non-integer values of $b$ have branching and
are 
discarded.  
  
We proceed to apply Painlev\'{e} analysis to
(\ref{eq:ode}) 
for integer $b\neq 0,1$, seeking leading order
behaviour 
at a movable point $z_0$ of the form  
$p\sim a(z-z_0)^\mu$ for integer exponent $\mu$. For
all 
integers $b\leq -2$ the only possible balance is 
$\mu=4/(3-b)$, which is non-integer and hence gives
algebraic 
branching. In the special case $b=-1$ there are four
possible 
balances with $\mu=1$, with $a$ and the resonances
depending 
on the value of $c$; we have checked that no value of 
$c$ gives all integer resonances, so the Painlev\'{e}
test 
is failed. For the remaining cases of integer $b\geq
2$ 
we find $\mu=1$ with $a^2=1$ or $a^2=4$. For 
integer $b\geq 4$ there is also 
the balance $\mu=4/(3-b)$ which is in general
non-integer,  
except for the special cases $\mu=-4,-2,-1$ 
for $b=4,5,7$ respectively. Thus all integer values of
$b$  
apart from $b=2,3,4,5,7$ 
are ruled out by the (strong) Painlev\'{e} test due to
algebraic branching; but they could still be analysed
by 
the weak Painlev\'{e} test if we allow such branching.

Let us consider in more detail 
the first two types of balance for 
integer $b\geq 2$. 
When $p\sim\pm (z-z_0)$ we have a non-principal
balance  
with resonances $r=-1,-1,3$. Interestingly, the
resonance 
condition at 
$r=3$ is failed when $b=2$, the obstruction being the 
$c^{-1}$ term (so for these balances 
the test is only passed in the limit 
$c\to\infty$), but satisfied for all integer $b\geq
3$. 
However, for the principal 
balances $p\sim\pm 2(z-z_0)$ the resonances are 
$r=-1,1/2,3/2$ which means there is algebraic  
branching and so the (strong) 
Painlev\'{e} test is failed for any $b$. We have
further  
checked whether the weak Painlev\'{e} test of
\cite{weak} 
could be satisfied by allowing an expansion in powers 
of $(z-z_0)^{1/2}$ in the principal balance. 
However, the resonance condition is satisfied at
$r=1/2$ 
but failed at $r=3/2$, meaning that this 
expansion with square root branching cannot 
represent the general solution as it doesn't contain 
enough arbitrary constants. The arbitrariness 
can only be restored 
by adding infinitely many terms in powers 
of $\log (z-z_0)$, and so no form of Painlev\'{e}
property 
can be recovered.  
The existence of logarithmic branching in both the 
principal and non-principal balances is a strong 
indication of non-integrability. 

It is interesting to observe that when the first term 
on the right hand side of (\ref{eq:ode}) is absent 
(the limit $c\to\infty$), it admits exact solutions 
in terms of trigonometric/hyperbolic functions, 
corresponding to the first order reductions   
$$ 
(p')^2=1+2dp+\frac{1}{3}(8e-d^2)p^2, \qquad 
(p')^2=4+8dp+\frac{8}{3}(2d^2-e)p^2.  
$$ 

In fact we can also see that the original equation
(\ref{eq:fifth}) 
fails 
the weak Painlev\'{e} test directly. For the
Camassa-Holm equation 
(\ref{eq:ch}) the test is satisfied by a principal
balance 
$$ 
u\sim -\phi_t/\phi_x + a\phi^{\frac{2}{3}}+\ldots 
$$ 
with resonances $-1,0,2/3$, with the singular manifold
$\phi (x,t)$ 
and $a(x,t)$ being arbitrary. For (\ref{eq:fifth})
there is an 
analogous balance 
$$ 
u\sim -\phi_t/\phi_x + a\phi^{\frac{4}{3}}+\ldots,  
$$   
with resonances $-1,0,4/3,(1\pm\sqrt{41})/6$; the
presence of 
irrational resonances implies logarithmic branching. 
 
\section{Prolongation algebra method} 

\setcounter{equation}{0}

While Painlev\'{e} analysis is a good heuristic tool
for isolating 
potentially integrable equations, it can never be said
to provide 
definite proof of non-integrability. If one gives a  
precise definition of integrability in terms of
existence of infinitely 
many commuting symmetries, then the symmetry approach
of Shabat et al 
\cite{shabat} gives necessary conditions for
integrability (but does  
not provide a constructive way to find a Lax pair or
linearization 
when such conditions are satisfied). The symmetry
approach has only 
very recently been extended \cite{mik} so that it can
be 
applied to nonlocal or non-evolution equations 
such as (\ref{eq:fifth}), (\ref{eq:ch}). As an
alternative, we 
apply the prolongation algebra method of Wahlquist and
Estabrook 
\cite{we}, and directly seek a Lax pair for
(\ref{eq:fifth}) 
in the form of a compatible 
linear system 
\beq 
\Psi_x=U\Psi, \qquad \Psi_t=V\Psi 
\label{eq:laxpair} 
\eeq 
for suitable matrices $U,V$ (usually taking values in
the 
fundamental representation of a semi-simple Lie
algebra) 
which should depend on $u$ and its 
derivatives, and on a spectral parameter. We have
found the 
clear presentation of the method in \cite{fordy} very
useful. 

The compatibility of the system (\ref{eq:laxpair})
yields 
the zero curvature equation 
\beq 
U_t-V_x+[U,V]=0, 
\label{eq:zeroc} 
\eeq 
and the essence of the Wahlquist-Estabrook method is
that given 
the original PDE (in this case (\ref{eq:fifth})) one 
may use (\ref{eq:zeroc}) to derive the functional
dependence 
of $U,V$ on $u,u_x,$ etc. A negative result means that
no Lax pair 
of a suitable form exists, suggesting that the
equation   
is not integrable, but of course this is sensitive to
the initial 
assumptions that are made on the functional form of
$U,V$.   

For ease of notation we will denote the $n$th
derivative $u_{nx}=u_n$. 
Given that (\ref{eq:fifth}) can be written as a
conservation law 
for $m$ as in  
(\ref{eq:fif}), a reasonable ansatz is to assume that 
$$ 
U=U(m), \qquad V=V(u,u_1,u_2,u_3,u_4)  
$$ 
(with dependence on the spectral parameter
suppressed).  
Given the known form of the zero curvature
representations for 
the equations (\ref{eq:ch}), (\ref{eq:td}) 
we further assume that $U$ is linear in $m$, so that 
$$ 
U=Am+B\equiv (u_4-5u_2+4u)A+B, 
$$ 
where $A,B$ are constant matrices (independent of
$x,t$, but 
potentially dependent on the spectral parameter). 
Substituting this ansatz into (\ref{eq:zeroc}), and
using 
(\ref{eq:fif}) to eliminate the $t$ derivative $m_t$,
we find 
$$ 
(-uu_5-2u_1u_4+5uu_3+10u_1u_2-12uu_1)A-u_5 V_{u_4} 
-u_4 V_{u_3}-u_3 V_{u_2}-u_2 V_{u_1} 
$$ 
\beq
-u_1 V_{u} 
+(u_4-5u_2+4u)[A,V]+[B,V]=0  
\label{eq:start} 
\eeq 
(with subscripts on $V$ denoting partial derivatives).
None  
of the matrices depend on $u_5$, so (\ref{eq:start})
is linear 
in $u_5$. In particular, the coefficient of $u_5$ must
vanish, 
giving the equation $V_{u_4}=-uA$ which integrates
immediately to 
yield 
\beq 
V=-uu_4 A+\Gamma(u,u_1,u_2,u_3), 
\label{eq:vres} 
\eeq 
where $\Gamma$ is so far arbitrary and must be
determined from 
the remaining terms in (\ref{eq:start}). 

At the next step we substitute for $V$ in the rest of
(\ref{eq:start}) 
to obtain 
$$ 
(-u_1u_4+5uu_3+10u_1u_2-12uu_1)A 
-u_4\Gamma_{u_3}-u_3\Gamma_{u_2}-u_2\Gamma_{u_1}-u_1\Gamma_{u}
$$          
\beq 
+(u_4-5u_2+4u)[A,\Gamma]+uu_4[A,B]+[B,\Gamma]=0. 
\label{eq:step1} 
\eeq 
The coefficient of $u_4$ gives the equation 
$$ 
\Gamma_{u_3}=[A,\Gamma+uB]-u_1A 
$$ 
which can be integrated exactly as 
\beq 
\Gamma=e^{u_3A}\Delta(u,u_1,u_2)e^{-u_3A}-u_1u_3A-uB, 
\label{eq:gamres} 
\eeq 
where $\Delta$ is the arbitrary function of
integration. From 
(\ref{eq:gamres}) we see the presence of 
${Ad}\,\exp u_3A= \exp ({ad}\, u_3A)$ acting on
$\Delta$, which 
would imply exponential-type dependence on $u_3$ in
the Lax pair 
unless $({ad}\, u_3A)^n \Delta=0$ for some positive
integer $n$. 
Such exponential dependence would seem unlikely given
that the 
original equation (\ref{eq:fifth}) is polynomial in
$u$ and its 
derivatives, and we will seek assumptions that
prohibit infinitely 
many non-zero commutators occurring in
(\ref{eq:gamres}).      

Substituting for $\Gamma$ from (\ref{eq:gamres}) in
the 
$u_4$-independent terms of (\ref{eq:step1}) and
applying 
${Ad}\,\exp (-u_3A)$ we get 
$$ 
(u_2u_3+5uu_3+10u_1u_2-12uu_1)A  
-u_3\Delta_{u_2}  
-u_2\Delta_{u_1}-u_1\Delta_{u} 
+(-5u_2+4u)[A,\Delta]  
$$ 
\beq 
+[e^{-u_3A}Be^{u_3A},\Delta]+(u_1u_3+5uu_2-4u^2)e^{-u_3A}Ce^{u_3A}=0,
\label{eq:step2} 
\eeq 
where we have set 
$$ 
C=[A,B]. 
$$  
Potentially (\ref{eq:step2}) is an infinite power
series in $u_3$,  
each coefficient of which must vanish. The simplest
assumption 
we can make to terminate the series is to take 
\beq 
[A,C]=0  
\label{eq:com} 
\eeq 
which  implies 
$$ 
{Ad}\, e^{-u_3A}(B)=e^{{ad}(-u_3A)}(B)=B-u_3C, 
\quad {Ad}\, e^{-u_3A}(C)=e^{{ad}(-u_3A)}(C)=C, 
$$   
and hence (\ref{eq:step2}) becomes linear in $u_3$. A 
fortunate consequence of (\ref{eq:com}) is that the
coefficient of 
$u_3$ gives 
$$ 
\Delta_{u_2}=(u_2+5u)A-[C,\Delta]+u_1C, 
$$ 
which integrates exactly without further assumption  
to yield 
\beq 
\Delta=e^{-u_2C}E(u,u_1)e^{u_2C}+ 
\left(\frac{1}{2}u_2^2+5uu_2\right)A+u_1u_2C.
\label{eq:delres} 
\eeq 

The remaining terms in (\ref{eq:step2}), after acting
with 
$Ad \,\exp u_2C$,  now become 
$$ 
(5u_1u_2-12uu_1)A-\left(\frac{3}{2}u_2^2+4u^2\right)C 
-u_2E_{u_1}-u_1E_{u}+(-5u_2+4u)[A,E] 
$$ 
\beq 
+[e^{u_2C}Be^{-u_2C},E]+u_1u_2[e^{u_2C}Be^{-u_2C},C]=0.
\label{eq:step3} 
\eeq 
Once again we are faced with an infinite power series,
this time in 
$u_2$. Before looking for further simplifying
assumptions, 
we note that the coefficient of the term linear in
$u_2$ is just 
$$ 
-E_{u_1}+u_1(5A+[B,C])-[5A+[B,C],E]=0, 
$$ 
which integrates immediately to 
\beq 
E=e^{-u_1D}Z(u)e^{u_1D}+\frac{1}{2}u_1^2D, \qquad
D=5A+[B,C]. 
\label{eq:eres} 
\eeq 
To analyse the other terms in (\ref{eq:step3}) we find
it convenient 
to introduce the quantities 
$$ 
F=[C,[C,B]], \qquad G=[D,B], 
$$ 
and note that the identities 
\beq 
[A,D]=0=[A,F], \qquad [A,G]=-[C,D]=F, \qquad
[[B,C],D]=0  
\label{eq:comm1} 
\eeq 
all hold. 

The coefficient of $u_2^2$ in (\ref{eq:step3}) is then

\beq 
-\frac{3}{2}C+\frac{1}{2}[F,E]-u_1 F=0, 
\label{eq:quad} 
\eeq 
and the coefficient of $u_2^0$ gives 
$$ 
-12uu_1A-4u^2e^{u_1D}Ce^{-u_1D}-u_1Z_u 
$$ 
\beq 
+4u[A,Z] 
+\left[e^{u_1D}Be^{-u_1D},
Z+\frac{1}{2}u_1^2D\right]=0 
\label{eq:step4} 
\eeq 
(after substituting for $E$ from (\ref{eq:eres}) and 
acting with $Ad\, e^{u_1D}$). We shall not need to
consider 
the equations $[(ad\, C)^n B,E]-nu_1(ad\, C)^n B=0$
occurring at 
$u_2^n$, $n\geq 3$. Instead we look at the coefficient
of $u_1$ in 
(\ref{eq:step4}), which is 
\beq 
-Z_u+[G,Z]-4u^2F-12uA=0. 
\label{eq:zeq} 
\eeq 
We are unable to integrate this directly without
making a further 
assumption, the simplest possible being 
\beq 
[F,G]=0 
\label{eq:comm2} 
\eeq 
which implies 
\beq 
Z=e^{Gu}\Theta e^{-Gu}+\frac{2}{3}u^3 F-6u^2 A. 
\label{eq:zres} 
\eeq 
We must now use the remaining equations to determine
the commutation 
relations for the constant Lie algebra elements
$A,B,C,D,F,G,\Theta$. 

Looking at the coefficient of $u_1^0$ in
(\ref{eq:quad}) we see that 
$$ 
-\frac{3}{2}C+\frac{1}{2}[F,Z]=0, 
$$ 
and using (\ref{eq:comm1}) and  (\ref{eq:comm2}) with
(\ref{eq:zres}) 
yields      
$$ 
[F,\Theta]=3e^{-Gu}Ce^{Gu}, 
$$ 
which immediately implies  
\beq 
[F,\Theta]=3C, \qquad [C,G]=0. \label{eq:comm3} 
\eeq 
Using the Jacobi identity we also have 
$$ 
0=[C,G]=[C,[D,B]]=-[B,[C,D]]-[D,[B,C]] 
=[B,F], 
$$ 
which by further applications of the Jacobi identity
gives 
\beq 
[C,F]=0=[D,F]. \label{eq:comm4} 
\eeq 
Now we return to the equation (\ref{eq:quad}) and use
(\ref{eq:eres}) 
to evaluate the coefficient of $u_1$ as 
\beq  
-\frac{1}{2}[F,[D,Z]]-F=0. \label{eq:qual} 
\eeq     
Substituting for $Z$ as in (\ref{eq:zres}) and taking
the 
constant coefficient $u^0$ in (\ref{eq:qual}) we  
use (\ref{eq:comm1}), (\ref{eq:comm3}),
(\ref{eq:comm4}) to find 
$$ 
0=-\frac{1}{2}[F,[D,\Theta]]-F=\frac{1}{2}\Big( 
[\Theta,[F,D]]+[D,[\Theta,F]]\Big)-F  
$$ 
$$ 
= -\frac{3}{2}[D,C]-F= 
-\frac{5}{2}F. 
$$ 
   
Then $F=0$ implies $C=0$ from the first equation in
(\ref{eq:comm3}), 
and it is straightforward to show that the Lax pair
(\ref{eq:laxpair}) 
collapses down to the trivial case $[U,V]=0$, with
(\ref{eq:zeroc}) 
reducing to the scalar equation $m_t=\mathcal{F}_x$ as
in 
(\ref{eq:fif}).

\section{Conclusions} 

Since the fifth order PDE (\ref{eq:fifth}) is in the 
class of pulson equations studied by 
Fringer and Holm \cite{fringer} it admits exact
solutions   
in the form of a direct superposition of an arbitrary
number 
of pulsons (as in Figure
\ref{Gauss_InitialCond-waterfall/contour_fig}). 
These particular $N$-pulson solutions take the precise
form (\ref{eq:pulsons}), and by the general results 
of \cite{fringer} we know that for $N=2$ the 
Hamiltonian equations (\ref{eq:geo})  
describing the two-body dynamics  
are integrable. However, several different
considerations 
provide strong evidence that the fifth order PDE
(\ref{eq:fifth}) 
is not integrable in the sense of admitting a Lax pair
and 
being solvable by the inverse scattering transform.  

It is well known that 
integrable PDEs such as the Korteweg-deVries equation 
or Camassa-Holm \cite{ch, ch2}  admit 
a compatible pair of Hamiltonian structures which 
together define a recursion operator generating
infinitely 
many higher symmetries. We have tried unsuccessfully 
to find an analogous 
bi-Hamiltonian or Lagrangian formulation for the 
fifth order equation (\ref{eq:fifth}), but as far 
as we are aware it admits only the single Hamiltonian 
structure (\ref{eq:lph}).

Both the Camassa-Holm equation (\ref{eq:ch}) and the 
new integrable equation (\ref{eq:td}) isolated by 
Degasperis and Procesi \cite{dega} exhibit the weak
Painlev\'{e} 
property of \cite{weak}, with algebraic branching in
local  
expansions representing a general solution. There are
many examples 
of Liouville integrable systems in finite dimensions
\cite{abenda} 
and Lax integrable PDEs \cite{hone} with this
property.  
For evolution equations transformations of 
hodograph type can restore the strong 
Painlev\'{e}
property \cite{hodo, hone}, and similarly in
\cite{needs, dhh} 
we have  used reciprocal transformations for the
non-evolutionary 
equations (\ref{eq:ch}), (\ref{eq:td}).  
In order to apply Painlev\'{e} analysis 
to (\ref{eq:fifth}) we have found it convenient to
employ a 
reciprocal transformation which removes the branching
at 
leading order, but further analysis of the travelling
wave 
reduction shows that 
there is still algebraic branching 
in the principal balances due to 
half-integer resonances. Furthermore, in both
principal and 
non-principal balances a resonance condition is
failed, and 
so after the transformation even the weak Painlev\'{e}
test  cannot be satisfied. 

We have also applied an integrability test which is 
perhaps  
less fashionable nowadays, namely 
the prolongation algebra method of Wahlquist and
Estabrook \cite{we}. 
By making certain simple assumptions we find that no
polynomial Lax 
pair 
of a suitable form exists for (\ref{eq:fifth}). 

We cannot expect the $N$-body pulson dynamical system
to be 
integrable for arbitrary $N>2$, since this would imply
the existence of an 
infinite integrable subsector within a non-integrable 
PDE. However, it would be good to find an analytical 
explanation for the apparent soliton-like behaviour 
of the pulson solutions, and their numerical stability
as evidenced by Figure
\ref{rear_collision-waterfall/contour_fig}.   
Further analytical and numerical studies will be
required to  
understand the stability properties of the pulson
solutions.    

\noindent {\bf Acknowledgements: } We are grateful to Martin Staley for 
providing the Figures. AH would like to thank the CR Barber Trust (Institute 
of Physics), the IMS (University of Kent) and the organizers of NEEDS 2002 for 
providing financial support.   
The authors are also grateful for the hospitality   
of the Mathematics Research Centre at the University of Warwick during the 
workshop Geometry, Symmetry and Mechanics II, where this work was 
completed.

\end{document}